\documentstyle[aps,prc]{revtex}

\begin{document}
\draft
\title{Isospin effects on rotational flow in intermediate energy heavy ion
collisions}
\author{Lie-Wen Chen,$^{1,2,3}$\/ Feng-Shou Zhang,$^{1,2,5}$ Zhi-Yuan Zhu$^{1,4}$}
\address{$^1$ Center of Theoretical Nuclear Physics, National Laboratory of Heavy Ion%
\\
Accelerator, Lanzhou 730000, China\\
$^2$ Institute of Modern Physics, Academia Sinica, P.O. Box 31, Lanzhou
730000, China\\
$^3$ Department of Applied Physics, Shanghai Jiao Tong University, Shanghai
200030, China\\
$^4$ Shanghai Institute of Nuclear Research, Academia Sinica, Shanghai
201800, China\\
$^5$ CCAST (World Laboratory), P.O. Box 8730, Beijing 100080, China}
\maketitle

\begin{abstract}
Within the framework of an isospin-dependent quantum molecular dynamics
model, the rotational flow in reactions of $^{58}$Fe + $^{58}$Fe and $^{58}$%
Ni + $^{58}$Ni at 40 MeV/nucleon for different impact parameters is
investigated by analyzing the mid-rapidity azimuthal distribution. The
rotational observables are also calculated semiquantitatively. For the first
time, it is found that the more neutron-rich system ($^{58}$Fe + $^{58}$Fe)
exhibits stronger rotational collective flow. This isospin dependence of
rotational collective flow is more appreciable in semi-peripheral collisions
and it is shown to mainly result from the isospin dependence of
nucleon-nucleon cross section rather than the symmetry energy. Meanwhile, it
is indicated that the rotational flow depends strongly on the impact
parameter.
\end{abstract}


The azimuthal distribution in heavy ion collisions (HIC's) has proved very
useful in the study of nuclear equation of state (EOS) as well as reaction
dynamics because of its sensitivity to collective flow\cite{TsangPLB84}. In
particular, the rotational flow may still exist even though the directed
flow has disappeared\cite{SoffPRC95}. With the recent advance in radioactive
nuclear beam physics, one can investigate isospin degrees of freedom in
nuclear reactions at wide energy ranges for different projectile-target
combinations\cite{LbaIJMP98,ClwPRC98}. Therefore, it is very significant to
explore the isospin effects on the azimuthal distribution through studying
HIC's induced by the radioactive nuclei since it may provide people a better
probe of isospin dependent reaction dynamics.

We report here results of the first theoretical study on the isospin
dependence of the azimuthal asymmetry rotational flow from reactions of $%
^{58}$Fe + $^{58}$Fe and $^{58}$Ni + $^{58}$Ni at 40 MeV/nucleon for
different impact parameters within the framework of an isospin dependent
quantum molecular dynamics (IQMD) model which includes the symmetry energy,
Coulomb interaction, isospin-dependent experimental $N$-$N$ cross sections,
and particularly the isospin dependent Pauli blocking\cite{ClwPRC98,ClwJPG97}%
. In the present calculations, the so-called soft EOS with an
incompressibility of $K=200$ MeV is used and the symmetry strength $C=32$
MeV without particular consideration\cite{AichelinPR91}. In the
initialization process of the IQMD model, the neutron and proton are
distinguished from each other\cite{ClwPRC98} and meanwhile the nonphysical
rotations in the initialized nuclei have been removed.

In the IQMD model simulation, the reaction plane is known $a\ priori$ and it
is defined as the $x$-$z$ plane ($z$-axis corresponds to the beam
direction). The azimuthal angle with respect to the reaction plane can be
written as

\begin{equation}
\phi =\arctan (P_y/P_x).
\end{equation}
In the present calculations, the calculated results have adopted the average
values from 200 to 300 fm/c to accumulate the numerical statistics since the
azimuthal distributions have been well stable after 200 fm/c and 200 events
are simulated for each impact parameter. Fig. 1 displays the IQMD model
predicted normalized azimuthal distribution $dN/d\phi $ for all nucleons
from $^{58}$Fe + $^{58}$Fe (solid circles) and $^{58}$Ni + $^{58}$Ni (open
circles) at 40 MeV/nucleon and impact parameter $b=6$ fm. Figures 1 (a), (b)
and (c) correspond to the target-like rapidity region, mid-rapidity region
and projectile-like rapidity region, namely, the values of reduced
center-of-mass (c.m.) rapidity (y/y$_{\text{proj}}$)$_{\text{c.m.}}$ of
nucleons considered belong to $[-1.5,-0.5]$, $[-0.5,0.5]$ and $[0.5,1.5]$,
respectively. The errors shown are statistical. Also included in Fig. 1 are
the Legendre polynomial fits up to the second order for the resulting
azimuthal distribution, i.e.,

\begin{equation}
dN/d\phi =c(1+a_1\cos (\phi )+a_2\cos (2\phi )).
\end{equation}
The solid (dashed) line is the result of fit for $^{58}$Fe + $^{58}$Fe ($%
^{58}$Ni + $^{58}$Ni). In Eq. (2) the coefficient $a_1$ represents the
strength of the directed flow with preferred emission at $\phi =\pm
180^{\circ }$ (i.e., $a_1<0$) for low rapidity values (in the backward
hemisphere) and $\phi =0^{\circ }$ (i.e., $a_1>0$) for high rapidity values
(in the forward hemisphere) if the incident energy is above the balance
energy. The coefficient $a_2$ reflects a flattening of the ellipsoid. A
negative value of $a_2$ (i.e., the azimuthal distribution peaks at $\phi
=\pm 90^{\circ }$, simultaneously) reflects the squeeze-out effects and a
positive value (i.e., the azimuthal distribution peaks at $\phi =0^{\circ }$
and $\pm 180^{\circ }$,simultaneously) the rotational collective motion. One
can see from Fig. 1 that for low (target-like) rapidity region the azimuthal
distribution only peaks at $\phi =0^{\circ }$ while it only peaks at $\pm
180^{\circ }$ for high (projectile-like) rapidity region, which imply that
these reaction systems display negative deflections (corresponding to
negative scattering angles) since the incident energy is below the balance
energy\cite{ClwPRC98}.

It is indicated in Fig. 1 (b) that the azimuthal distribution peaks at $\phi
=0^{\circ }$ and $\pm 180^{\circ }$ simultaneously for mid-rapidity
nucleons, which are indicative of a rotation-like behavior. The strength of
the rotational flow, $a_2$, equals to 0.20151 and 0.18047 for $^{58}$Fe + $%
^{58}$Fe and $^{58}$Ni + $^{58}$Ni, respectively. In order to observe the
impact parameter dependence of rotational flow and explore its isospin
dependence at different impact parameters, we show values of $a_2$ for
mid-rapidity nucleons in Fig. 2 as a function of impact parameter $b$ for $%
^{58}$Fe + $^{58}$Fe (solid circles) and $^{58}$Ni + $^{58}$Ni (open
circles) at 40 MeV/nucleon. The errors shown are statistical. A strong
isospin dependence is observed in Fig. 2, namely, the more neutron-rich
system $^{58}$Fe + $^{58}$Fe displays stronger collective flow than the
system $^{58}$Ni + $^{58}$Ni, especially in the semi-peripheral collisions ($%
b=6$ fm). In addition, one can see that the rotational flow increases with
increment of impact parameter $b$ from near central to peripheral collisions
and then decreases, and finally disappears in the most peripheral collisions
since at such large impact parameter the interaction between the projectile
and target is very weak.

In order to see more clearly the rotational flow and its isospin dependence,
one can calculate some characteristic physical quantities of the rotation
for mid-rapidity nucleons, such as the angular momentum, the moment of
inertia, the angular velocity and its rotational energy, with the help of
classical mechanics. In the present IQMD model calculations, we simply
consider the nucleon system as a rigid body at every time point and then the
angular momentum about the normal of the reaction plane ($y$-axis) can be
calculated as follows,

\begin{equation}
L_y=\sum L_y(i)=\sum (z_ip_{x_i}-x_ip_{z_i}),
\end{equation}
where $x_i$, $z_i$, $p_{x_i}$, $p_{z_i}$ are components on the reaction
plane of ${\bf r}_i$ and ${\bf p}_i$ in the c.m. system. The moment of
inertia about the $y$-axis can be written as

\begin{equation}
I_y=\sum I_y(i)=\sum m_i(r_i^2-y_i^2).
\end{equation}
Therefore, the angular velocity around the $y$-axis can be expressed as

\begin{equation}
\omega _y=L_y/I_y,
\end{equation}
and its collective rotational energy as

\begin{equation}
E_{rot}=\frac 12I_y\omega _y^2.
\end{equation}
Nevertheless, the semiquantative description above may provide us some
intuitive information about the in-plane collective rotational behavior
although it is classical and very rough for HIC's at intermediate energies.
Fig. 3 illustrates the time evolution of the absolute value of angular
momentum per nucleon $\left| L_y/A\right| $, moment of inertia per nucleon $%
I_y/A$, absolute value of angular velocity $\left| \omega _y\right| $, and
rotational energy per nucleon $E_{rot}/A$ from $^{58}$Fe + $^{58}$Fe (solid
circles) and $^{58}$Ni+$^{58}$Ni (open circles) at 40 MeV/nucleon and $b=4$
fm for mid-rapidity nucleons. The errors shown are statistical. Meanwhile,
also included in Fig. 3 (a) are the time evolution of the absolute value of
angular momentum per nucleon around the $y$-axis for the whole system of
reactions $^{58}$Fe+$^{58}$Fe (solid squares) and $^{58}$Ni+$^{58}$Ni (open
squares) and its amplified window figure for clarity, which indicate that
the conservation of angular momentum is destroyed slightly, namely, the
angular momentum per nucleon for $^{58}$Fe+$^{58}$Fe ($^{58}$Ni+$^{58}$Ni )
changes slightly, i.e., from 1.388 (1.388) $\hbar $/nucleon to 1.348 (1.343) 
$\hbar $/nucleon in the whole reaction process, and meanwhile the isospin
dependence and the statistical errors are very small. The values for $^{58}$%
Ni + $^{58}$Ni have been offset with 2 fm/$c$ in the horizontal direction
for clarity. In fact, both $L_y/A$ and $\omega _y$ are negative by Eqs. (3)
and (5) since the reactions display negative deflections. From Fig. 3 (a),
one can see clearly that the increment of angular momentum becomes more
slowly after about 125 fm/c and tends to its asymptotic value after a longer
time when the reaction system may have not any further interaction. The
moment of inertia reaches its minimum at about 40 fm/c when the reaction
systems reach their maximum overlap and then increases with the reaction
systems expanding. The angular velocity peaks at about 50 fm/c when the
system has just undergone the strongest exchanges of all kinds of degrees of
freedom and then becomes saturate after about 200 fm/c. Similarly, $%
E_{rot}/A $ has maximum at about 100 fm/c and then becomes saturate after
about 200 fm/c. From above analysis, one can conclude that there actually
exists in-plane collective rotational motion for mid-rapidity nucleons at
such incident energy and impact parameter. The similar results are observed
for other impact parameters. More importantly here, one can find from Fig. 3
that the more neutron-rich system $^{58}$Fe + $^{58}$Fe exhibits clearly
stronger rotational energy and angular momentum than system $^{58}$Ni + $%
^{58}$Ni, which confirms again the conclusion obtained in Fig. 2. Meanwhile,
these semiquantative calculations also support the view about extracting the
rotational flow from the azimuthal distribution analysis.

The isospin dependence of the rotational flow may be a result of the
competition among several mechanisms in the reaction dynamics, such as the
symmetry energy, isospin dependent {\sl N-N} cross sections, Coulomb energy,
the surface properties of the colliding nuclei, and so on. Here, we
investigate the influence of the symmetry energy and isospin dependent {\sl %
N-N} cross section on the rotational flow.

Using different symmetry energy strength $C$ and parametrizations of {\sl N-N%
} cross sections, we show in Fig. 4 the IQMD model predicted azimuthal
distribution from $^{58}$Fe + $^{58}$Fe (solid circles) and $^{58}$Ni + $%
^{58}$Ni (open circles) at 40 MeV/nucleon and $b=4$ fm for mid-rapidity
nucleons. The errors shown are statistical. Meanwhile, the results of
Legendre polynomial fits according to Eq. (2) for $^{58}$Fe + $^{58}$Fe
(solid line) and $^{58}$Ni + $^{58}$Ni (dashed line) as well as the
resulting $a_1$ and $a_2$ are also included in Fig. 4. In Fig. 4 (a) we use $%
C=0$ (no symmetry energy) and Cugnon's {\sl N-N} cross section $\sigma _{%
\text{Cug}}$ which is isospin independent. It is indicated that $a_2$ of $%
^{58}$Fe + $^{58}$Fe is slightly less than that of $^{58}$Ni + $^{58}$Ni,
which may result from the difference of Coulomb interaction and surface
properties between the two systems. The case of using $C=32$ MeV and $\sigma
_{\text{Cug}}$, is plotted in Fig. 4 (b). For the results shown in Fig. 4
(c) we use $C=32$ MeV and experimental {\sl N-N} cross section $\sigma _{%
\text{exp}}$ which is isospin dependent. One can see from Fig. 4 that the
values of $a_1$ are very small since the transverse momentum disappears at
mid-rapidity and meanwhile can find that the symmetry energy enhances the
rotational flow ($a_2$) while $\sigma _{\text{exp}}$ reduces it more
strongly. It is also indicated that the influence of $\sigma _{\text{exp}}$
on system $^{58}$Ni + $^{58}$Ni is stronger than that on the system $^{58}$%
Fe + $^{58}$Fe, which just results in the observed isospin dependence. This
is easy to understand since the neutron-proton cross section is about three
times larger than the neutron-neutron or proton-proton cross section for $%
\sigma _{\text{exp}}$, which results in more {\sl N-N} collisions for $^{58}$%
Ni + $^{58}$Ni. From above analysis, one can conclude that the isospin
dependence of rotational flow seems to mainly result from the isospin
dependence of {\sl N-N} cross section and the symmetry potential has little
influence on it.

In summary, by using the IQMD model, we studied for the first time the
in-plane rotational flow in reactions of $^{58}$Fe + $^{58}$Fe and $^{58}$Ni
+ $^{58}$Ni at 40 MeV/nucleon for different impact parameters by analyzing
the mid-rapidity azimuthal distribution and calculating semiquantitatively
the rotational observables. A strong isospin dependence of the rotational
flow has been found, namely, the more neutron-rich system exhibits stronger
rotational flow. This isospin dependence is more appreciable in
semi-peripheral collisions and it is shown to mainly result from the isospin
dependence of {\sl N-N} cross section rather than the symmetry energy.
Meanwhile, it is indicated that the rotational flow depends strongly on the
impact parameter, namely, it increases with increment of impact parameter
from near central to peripheral collisions and then decreases, and finally
disappears in the most peripheral collisions.

The authors would like to thank W. Q. Shen for interesting discussions. This
work was supported by the National Natural Science Foundation of China under
Grant NOs. 19609033, 19875068, and 19847002, and the Foundation of the
Chinese Academy of Sciences.

\section*{Figure captions}

\begin{description}
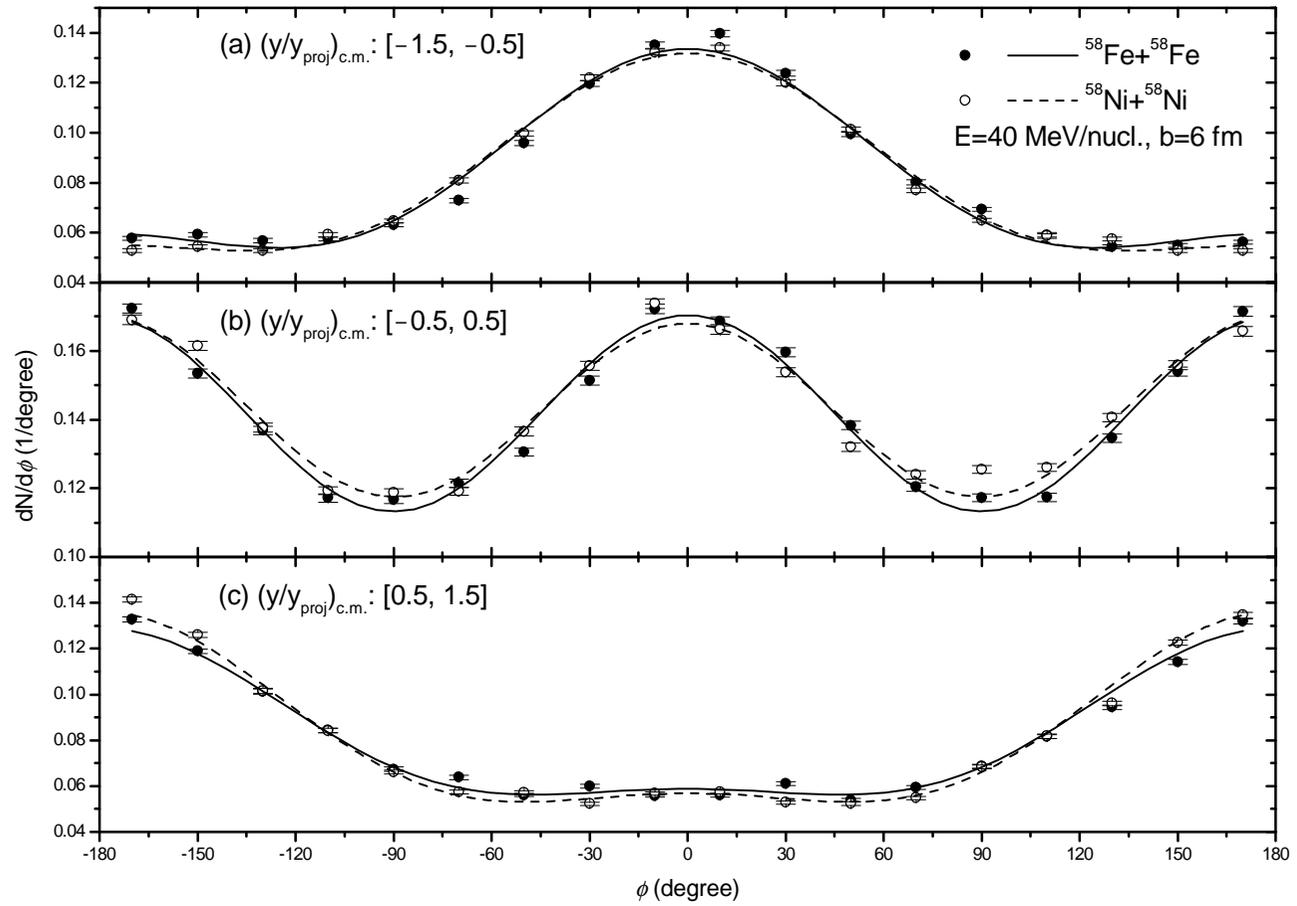
\item[Fig. 1]  The IQMD model predicted normalized azimuthal distribution $%
dN/d\phi $ for all nucleons from $^{58}$Fe + $^{58}$Fe (solid circles) and $%
^{58}$Ni + $^{58}$Ni (open circles) at 40 MeV/nucleon and impact parameter $%
b=6$ fm for different rapidity regions, i.e., the target-like rapidity
region (a), mid-rapidity region (b) and projectile-like rapidity region (c).
The solid (dashed) line represents the Legendre polynomial fit up to second
order for the calculated result of $^{58}$Fe + $^{58}$Fe ($^{58}$Ni + $^{58}$%
Ni).

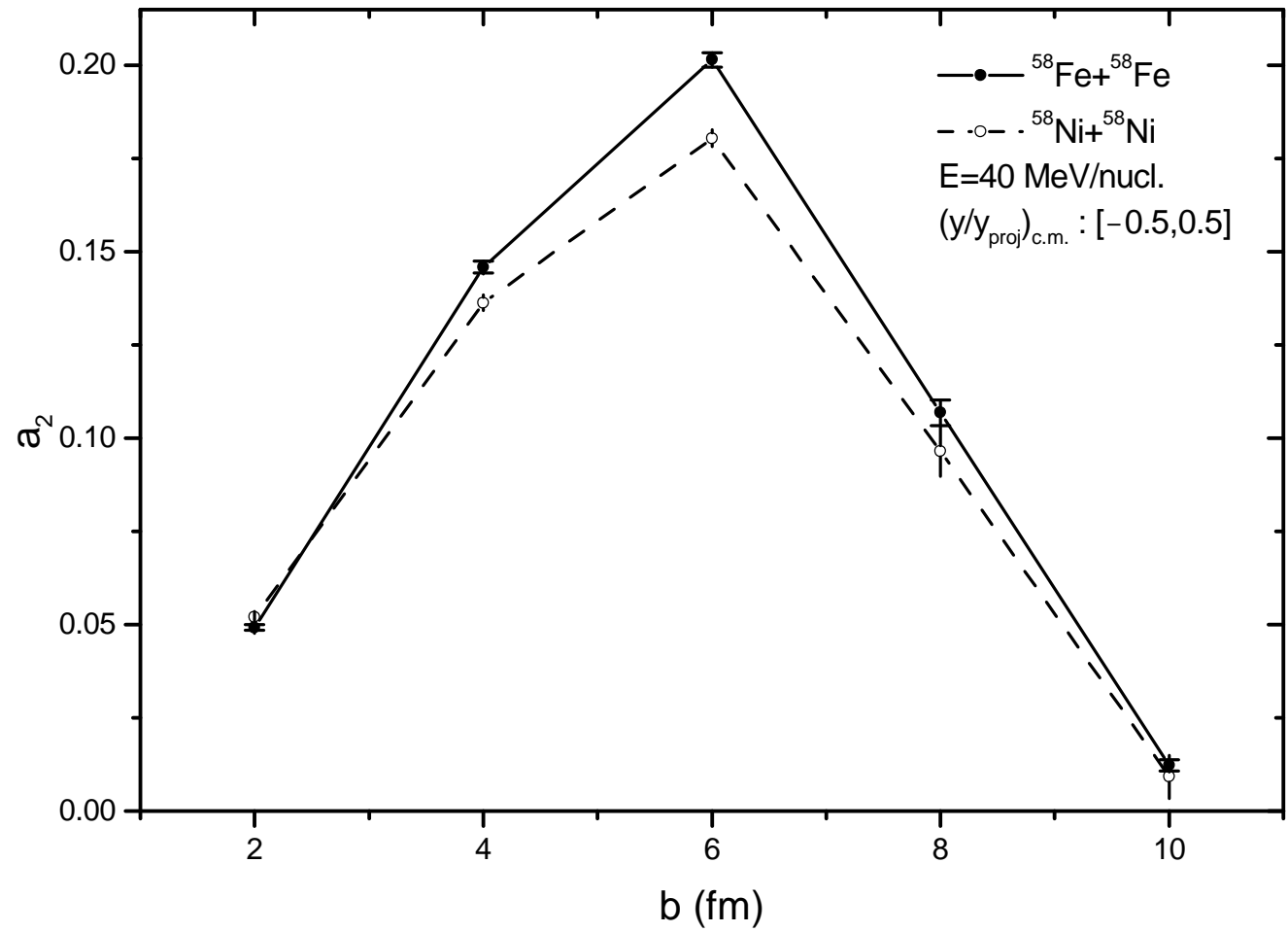
\item[Fig. 2]  The values of $a_2$ for mid-rapidity nucleons as a function
of impact parameter $b$ for $^{58}$Fe + $^{58}$Fe (solid circles) and $^{58}$%
Ni + $^{58}$Ni (open circles) at 40 MeV/nucleon. The lines are plotted to
guide the eye.

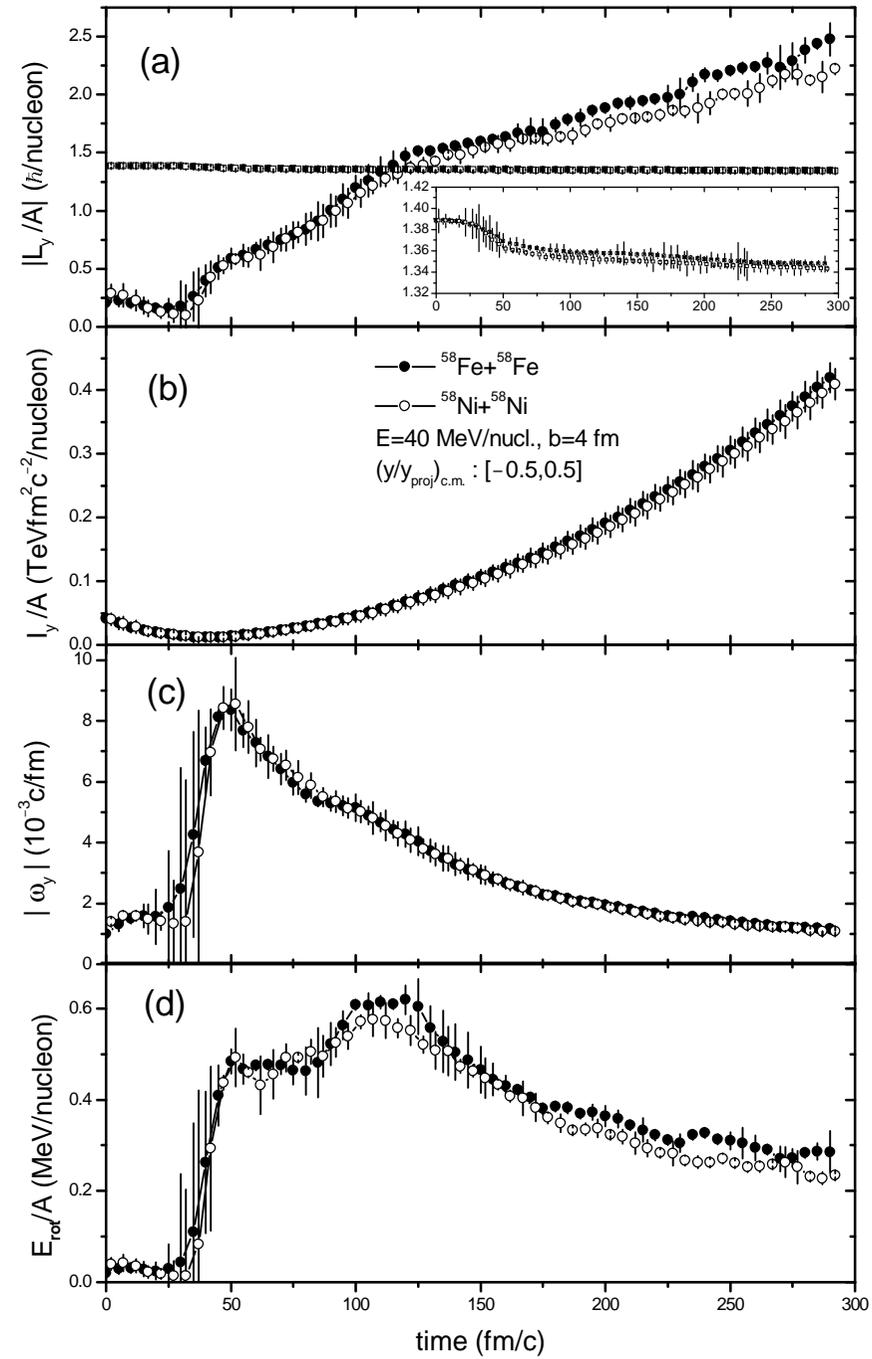
\item[Fig. 3]  The time evolution of the absolute value of angular momentum
per nucleon $\left| L_y/A\right| $ (a), moment of inertia per nucleon $I_y/A$
(b), absolute value of angular velocity $\left| \omega _y\right| $ (c), and
rotational energy per nucleon $E_{rot}/A$ (d) from $^{58}$Fe + $^{58}$Fe
(solid circles) and $^{58}$Ni + $^{58}$Ni (open circles) at 40 MeV/nucleon
and $b=4$ fm for mid-rapidity nucleons. Also included in (a) are the time
evolution of the absolute value of angular momentum per nucleon for the
whole system of reactions $^{58}$Fe+$^{58}$Fe (solid squares) and $^{58}$Ni+$%
^{58}$Ni (open squares) and its amplified window figure.

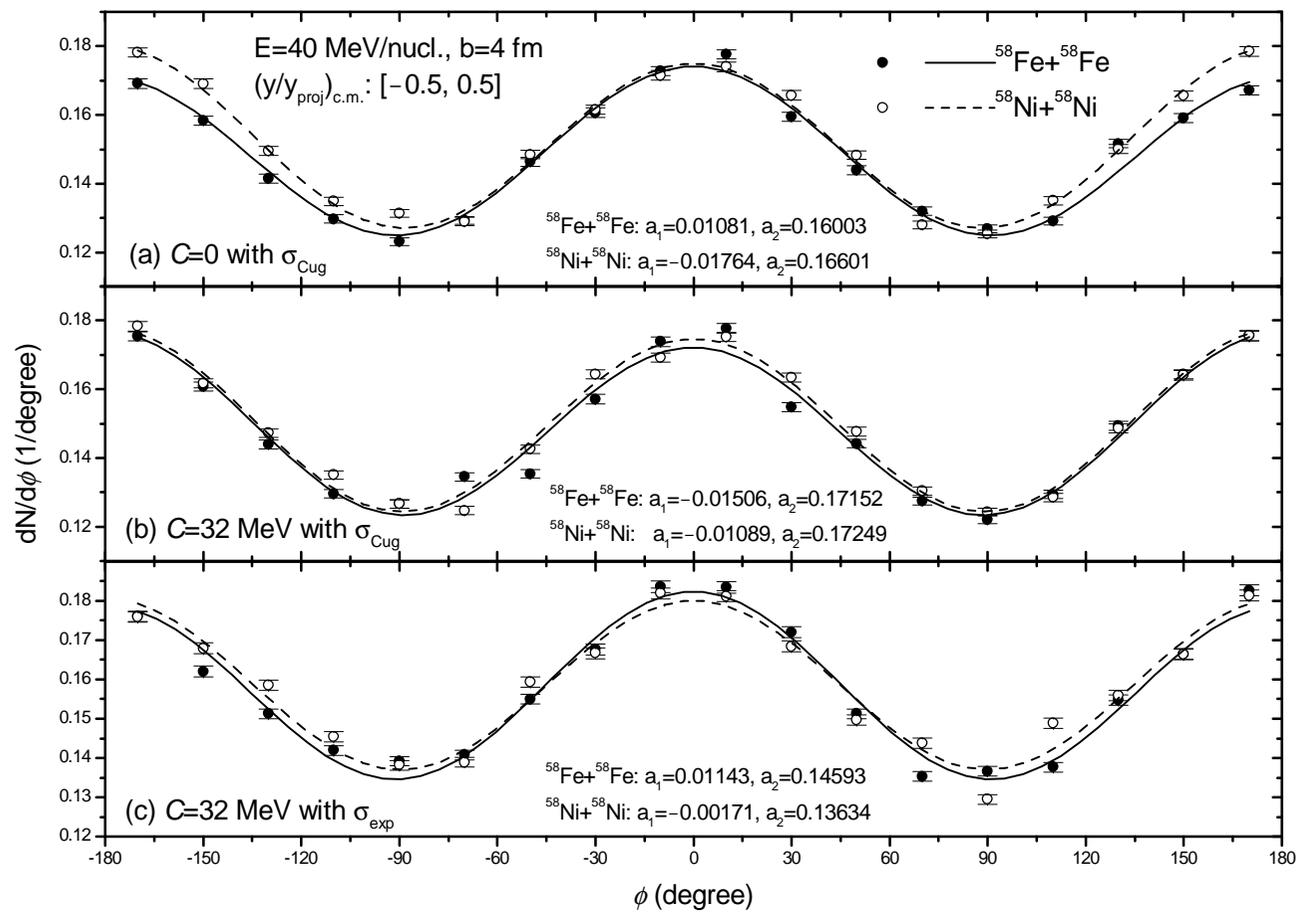
\item[Fig. 4]  The IQMD model predicted azimuthal distribution for $^{58}$Fe
+ $^{58}$Fe (solid circles) and $^{58}$Ni + $^{58}$Ni (open circles) at 40
MeV/nucleon and $b=4$ fm for mid-rapidity nucleons by using different
symmetry energy strength $C$ and parametrizations of {\sl N-N} cross
sections: $C=0$ (no symmetry energy) with Cugnon's {\sl N-N} cross section $%
\sigma _{\text{Cug}}$ (a), $C=32$ (MeV) with $\sigma _{\text{Cug}}$ (b), and 
$C=32$ (MeV) with experimental {\sl N-N} cross section $\sigma _{\text{exp}}$
(c). Meanwhile, the results of Legendre polynomial fits according to Eq. (2)
for $^{58}$Fe + $^{58}$Fe (solid line) and $^{58}$Ni + $^{58}$Ni (dashed
line) as well as the resulting $a_1$ and $a_2$ are also included.
\end{description}

\end{document}